\documentclass[12pt]{iopart}
\usepackage{graphicx}

\begin{document}

\title{{Ill-Behaved Convergence of a Model of the Gd$_3$Ga$_5$O$_{12}$ 
Garnet Antiferromagnet with Truncated Magnetic Dipole-Dipole Interactions}}

\author{Taras Yavors'kii$^1$, Michel J P Gingras$^{1,2}$, Matthew Enjalran$^3$}
\address{
$^1$Department of Physics and Astronomy, University of Waterloo, Ontario, N2L 3G1, Canada}
\address{
$^2$Department of Physics and Astronomy, University of Canterbury,
Private Bag 4800, Christchurch, New Zealand}
\address{
$^3$Department of Physics, Southern Connecticut State University, New Haven, CT 06515, USA}

\ead{tarasyk@galadriel.uwaterloo.ca}

\begin{abstract}
Previous studies have found that calculations which consider
long-range magnetic dipolar interactions truncated at
a finite cut-off distance $R_c$ predict spurious (unphysical)
long-range ordered phases for Ising and
Heisenberg systems on the pyrochlore lattice.
In this paper we show that, similar to these two cases,
calculations that use truncated dipolar interactions to model
the Gd$_3$Ga$_5$O$_{12}$ garnet antiferromagnet also predict
unphysical phases with 
incommensurate ordering wave vector
${\bf q}_{\rm ord}$ that is very 
sensitive to the dipolar cut-off distance $R_c$.
\end{abstract}



\section{Introduction}
There are currently many highly-frustrated magnetic materials being
experimentally studied where the magnetic species consist of a rare-earth
4f ion, such as Ho$^{3+}$, Dy$^{3+}$, Gd$^{3+}$ or Tb$^{3+}$,
which can have a large magnetic dipole moment.
Because of the large moment,
the long-range dipole-dipole interactions in
these systems are an important part of
the full spin Hamiltonian.
The role of dipolar interactions in highly frustrated magnetic
Ising systems has been systematically investigated for the three-dimensional
pyrochlore lattice of corner-sharing tetrahedra
\cite{Gingras-CJP,Melko,Isakov-PRL}.

In highly-frustrated Heisenberg
antiferromagnets of corner-sharing triangles or tetrahedra, 
any state with zero total magnetic moment on each elementary triangle or tetrahedron unit is
a classical ground state
\cite{Moessner-Chalker}.
There are an infinite number of such spin configurations 
and this is 
why these systems fail 
to develop
conventional magnetic order at nonzero temperature
\cite{Moessner-Chalker}.  
In cases where 
dipolar
interactions are somewhat weaker than 
nearest-neighbor exchange interactions, one
might have naively assumed that the long-range dipolar interactions can be truncated 
at a finite cut-off distance, 
$R_c$,
since the
nearest-neighbor exchange energetically controls and enforces the nearest-neighbor
correlations. Previous studies on the Ising spin ice pyrochlore
systems~\cite{Gingras-CJP,Melko,Isakov-PRL}
and
the Heisenberg pyrochlore antiferromagnet~\cite{Enjalran-condmat,Cepas}
have found that this naive expectation is erroneous and that truncating the
dipolar interactions at a finite cut-off distance $R_c$ leads to
spurious (unphysical) long-range ordered phases, and that it is crucial
to consider dipolar interactions to infinite distance
($R_c=\infty$). In this paper we
report a third example of a highly frustrated 
spin system 
which is very sensitive to the dipolar cut-off.
Specifically, we consider a Heisenberg model on
the three-dimensional garnet lattice structure
of corner-shared triangles.
Our work extends the study of a dipolar Ising version of the 
model on a garnet lattice~\cite{Yoshioka04} and is
relevant to the ultimate understanding of the nature of
the incommensurate spin-spin correlations that develop in the Gd$_3$Ga$_5$O$_{12}$ garnet
(GGG) antiferromagnet below a temperature of
500~mK~\cite{Petrenko-PRL,Petrenko-Physica}.

\section{Model and method}
In order to investigate the problem of an adequate treatment of the dipolar interactions 
in GGG, we consider below a minimal model Hamiltonian $H$ for it.
To best expose the physics of a truncated dipolar lattice sum,
we ignore the  
effects of the quantum nature 
of the Gd$^{3+}$ spins,
lattice disorder \cite{Petrenko-PRL},
exchange interactions beyond nearest neighbors 
\cite{Petrenko-PRL,Wolf,condmat0511403}
and possible single ion anisotropy~\cite{Rimai},
all of which are potentially important 
for a thorough quantitative understanding of GGG.
We 
describe the magnetic $\rm Gd^{3+}$ spins ${\bf S}$ as 
classical and isotropic 
$n=3$ component
(Heisenberg) vectors of the length 
$\sqrt{({\rm S}({\rm S}+1))}$ 
(${\rm S}=7/2$)
coupled by frustrated antiferromagnetic nearest-neighbor exchange 
and long range dipolar interactions, of strength $J=0.107$ K and 
$D=0.0457$ K, respectively \cite{Wolf,condmat0511403}:
\begin{equation}
\nonumber
H= \;J\; \sum_{<i,j>} {\bf S}_i\cdot{\bf S}_j\; +
\;D\; \sum_{i>j}
\frac{1}{r_{ij}^3}\, \left[ {\bf S}_i\cdot{\bf S}_j-3\,({\bf S}_i\cdot\hat{r}_{ij}) ({\bf S}_j
\cdot\hat{r}_{ij})\right ].
\label{Ham}
\end{equation}
In Eq.~(\ref{Ham}), $i, j$ span the sites of the GGG lattice 
(see Fig. 2 in  Ref.~\cite{Wolf} for the GGG lattice structure)
which are
separated by vectors 
${\bf r}_{ij} \equiv r_{ij} \hat{r}_{ij}$
of directions $\hat{r}_{ij}$; $\langle i, j \rangle$ 
denotes pairs of nearest neighbors. 
The size of the GGG conventional cubic cell is $a= 12.349$ \AA 
\cite{Petrenko-PRL}.

\begin{figure*}[ht]
\begin{picture}(450,390)
\put(0,200){\includegraphics[width=50mm]{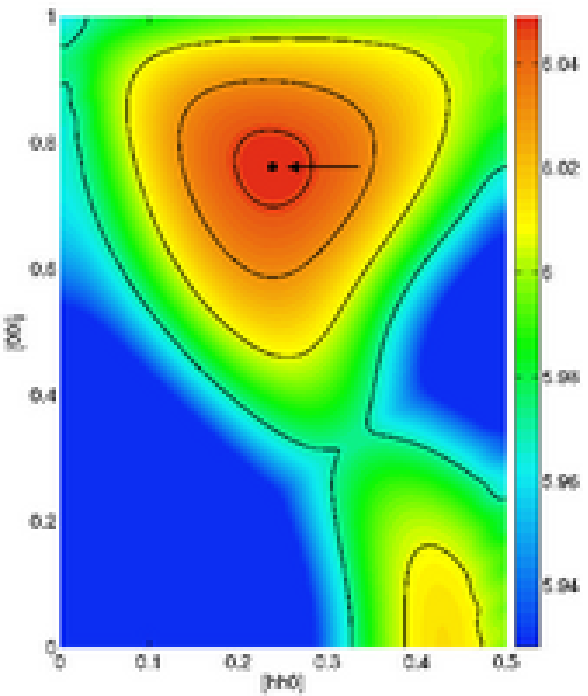}}
\put(150,200){\includegraphics[width=50mm]{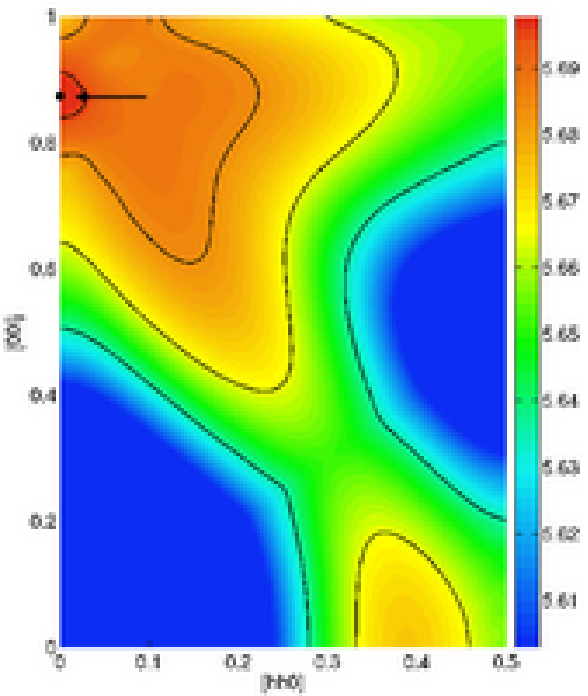}}
\put(300,200){\includegraphics[width=50mm]{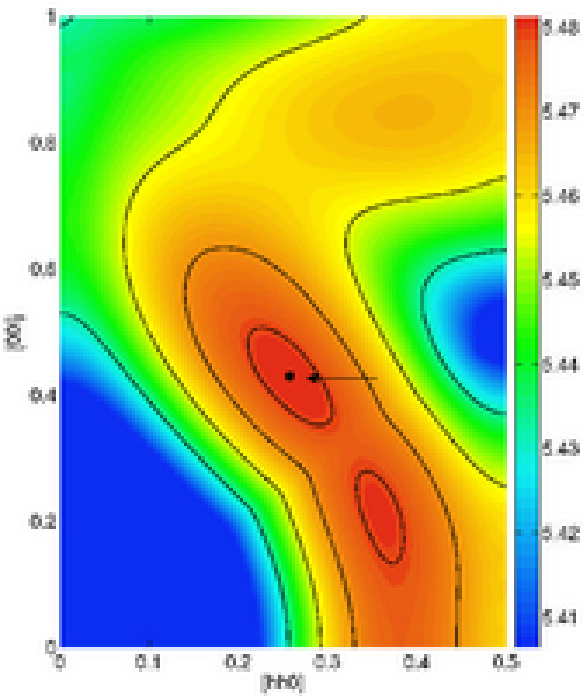}}
\put(0,15){\includegraphics[width=50mm]{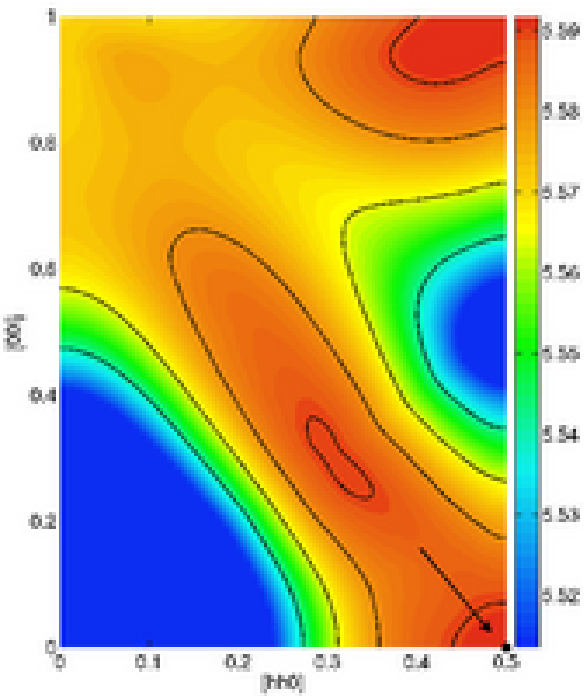}}
\put(150,15){\includegraphics[width=50mm]{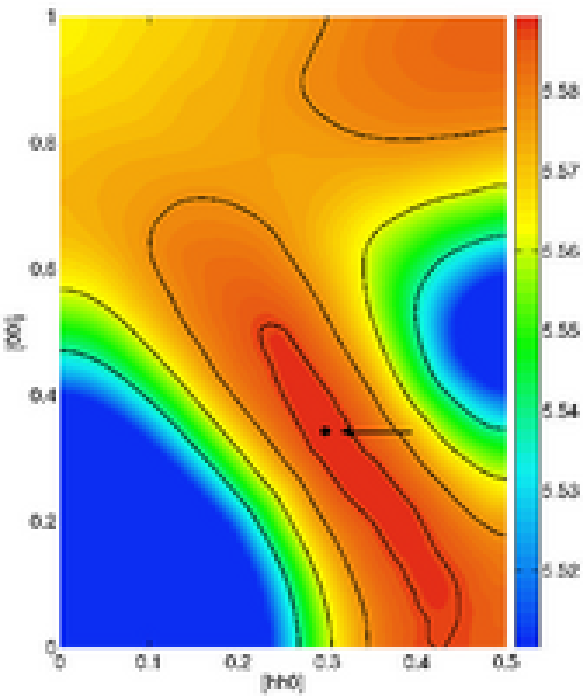}}
\put(300,15){\includegraphics[width=50mm]{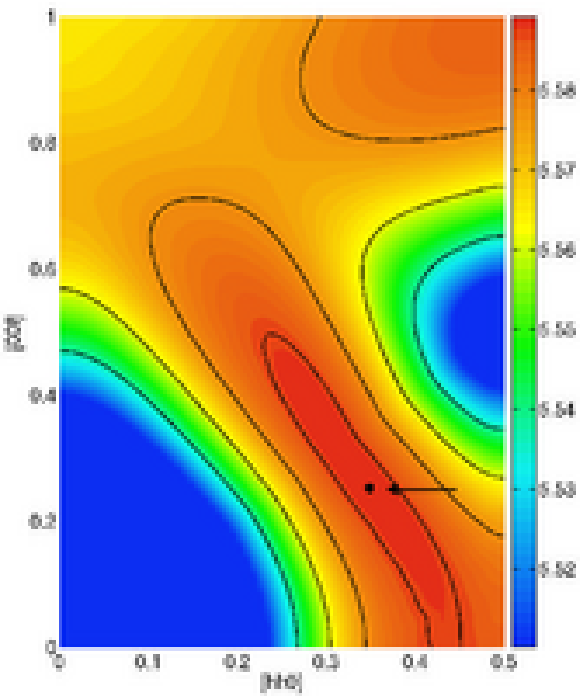}}
\end{picture}
\caption{\label{hhlRc.eps}
Each panel shows the upper branch
$\lambda^{\rm up}({\bf q})$ of eigenvalues of model (\ref{Ham})
as a function of wave vector ${\bf q}$ in the (hhl) plane at various 
cut-off distances $R_c$: $R_c=3,4,5$ (upper panels, left to right) and
$R_c=100,1000,\infty$ (lower panels).
The global maximum $\lambda_{\rm max}$
defines the ordering wave vector ${\bf q}_{\rm ord}$
(denoted by an arrow).
Its location is a complex function of $R_c$.
The isolines are drawn at 0.5\%, 3\%, 7\%, 13\%,  of the overall dispersion
downhill from the global maximum;
the actual values of the dispersions as well as of $\lambda_{\rm max}$
depend on $R_c$ (${\bf q}$ is measured in units of $2\pi/a$).
}
\end{figure*}

We aim to identify the critical (or, soft) 
modes of model (\ref{Ham}) for which a magnetic instability first
develops as the temperature, $T$, is reduced.
We consider
the soft mode spectrum in the Gaussian (mean-field theory, or MFT) approximation 
\cite{MFT}.
Following Ref.~\cite{MFT}, we apply MFT to calculate the neutron scattering intensity $I({\bf q})$.
This allows for a convenient way of analyzing the physical
influence of finite $R_c$ on the magnetic correlations as well as
for a direct comparison
with
experimental data~\cite{Petrenko-PRL}:
\begin{equation}
\label{Sofq}
I({\bf q}) = [f(|{\bf q}|)]^2 \times \lim_{N\rightarrow\infty}
1/N \sum_{ij} \langle {\bf S}_{i}^{\perp}
\cdot{{\bf S}}_{j}^{\perp} \rangle e^{\imath{\bf {q}}\cdot{\bf {r}}_{ij}}\,.
\end{equation}
Here, angular brackets denote a thermal average, $N$ is the number of spins,
${\bf S}_i^\perp$ represents the components of spin 
${\bf S}_i$ at site $i$ perpendicular
to the scattering vector ${\bf q}$, and $f(|{\bf q}|)$ is the
magnetic form-factor of Gd$^{3+}$\cite{Brown}.
The MFT expression for $I({\bf q})$ \cite{MFT} 
is obtained from the 
eigenvalues $\lambda^{\alpha}({\bf q})$ and eigenvectors of the
basis-diagonalization of the Fourier transform of 
exchange and dipolar interactions in Eq.~(\ref{Ham}) \cite{MFT}:
\begin{equation}
\label{IqMFT}
I({\bf q})  = [f(|{\bf q}|)]^2 \sum_{\alpha}
\frac{|{\bf F}^{\alpha}_{\perp}({\bf q})|^2}
{\left(n - \lambda^{\alpha}({\bf q})/T\right)} \,.
\end{equation}
Here $\alpha=\{1,\ldots,n\cdot N_b=36\}$ enumerates the eigenvalues and 
the vector ${\bf F}^{\alpha}_{\perp}({\bf q})$ incorporates information on the
eigenvectors and represents the role of the 
paramagnetic
form factor of the GGG
primitive unit cell, which contains $N_b=12$ ions.
The ordering wave vector ${\bf q}_{\rm ord}$  
is given by locating
in the first Brillouin zone the global maximum, $\lambda_{\rm max}$, 
of the maximum value (upper branch), $\lambda^{\rm   up} ({\bf q})$,
among the 36 $\lambda^\alpha ({\bf q})$ eigenvalues.
The mean field critical temperature,
$T_c^{\rm MFT}=\lambda_{\rm max}/n$, is used to define a
(positive) dimensionless temperature
$\tau=T/T_{\rm c}^{\rm MFT}-1$ to serve as a natural energy scale.

\section{Results}
With the aim of studying the effect of the dipolar cut-off 
on the magnetic correlations of model (\ref{Ham}),
we first compute, for various 
$R_c$, 
$\lambda^{\rm up}({\bf q})$
for arbitrary ${\bf q}$ in the first Brillouin zone.
Technically, we calculate the 
$\lambda^{\alpha}({\bf q})$
modes on a finite $32^3$ ${\bf q}$-space grid 
in the zone, and 
obtain their values at any
${\bf q}$ 
using a three dimensional cubic interpolation procedure. 
We verify the grid-independence of the results 
by considering denser grids and by cross-checking the interpolated results 
with exact calculations at judiciously chosen ${\bf q}$ values. 

We find that the dipolar term of Eq.~(\ref{Ham}) 
selects a unique ordering 
wave vector ${\bf q}_{\rm ord}$ 
with corresponding mode
$\lambda^{\rm up}({\bf q}_{\rm ord})$ 
out of the massively degenerate
spectrum of soft modes of the nearest-neighbor model at any cut-off distance $R_c>1$.
The ordering wavevector ${\bf q}_{\rm ord}$ was found to belong to the (hhl) planes of 
the first Brillouin zone for all $R_c>1$.
However, its location in those planes
is very sensitive to $R_c$.
We display this in Fig.~\ref{hhlRc.eps}
by showing the dependence of 
the 
$\lambda^{\rm up}({\bf q})$ 
modes on 
${\bf q}$ in the (hhl) plane
for $R_c=3,4,5,100,1000$. 
We note 
their
three important properties 
as reflected in Fig.~\ref{hhlRc.eps}. 
First, for each value of $R_c$, 
$\lambda^{\rm up}({\bf q})$ 
is characterized by a relatively small overall dispersion 
$\lambda_{\rm max}/\lambda_{\rm min}-1 \approx 10\%$ 
throughout the zone, where $\lambda_{\rm min}$ 
is the global minimum of the branch. 
Even within the interval 1\% below 
$\lambda_{\rm max}$ 
(or at 13\% of the overall dispersion, as delineated by the outermost isolines), 
$\lambda^{\rm up}({\bf q})$ 
covers the major part of the (hhl) plane.
Second, though the general topology of 
$\lambda^{\rm up}({\bf q})$ 
within the Brillouin zone 
is preserved with varying dipolar 
cut-off distance $R_c$, its fine structure is 
manifestly
sensitive to $R_c$.
The ordering wave vector ${\bf q}_{\rm ord}$ displays a non-monotonous
dependence upon $R_c$ as shown in Fig.~\ref{hhlRc.eps} by its movement throughout 
the zone.
Third, 
${\bf q}_{\rm ord}$ 
converges to a well-defined value, and so does the
dispersion of $\lambda^{\rm up}({\bf q})$
away from ${\bf q}_{\rm ord}$, at large 
values of $R_c$, as is evident
by comparing the $R_c=1000$ panel with the limiting case of $R_c=\infty$, 
recast by the Ewald method \cite{MFT}. These are the central results of the paper.

\begin{figure*}[ht]
\begin{picture}(450,330)
\put(5,320){\includegraphics[width=56mm,angle=-90]{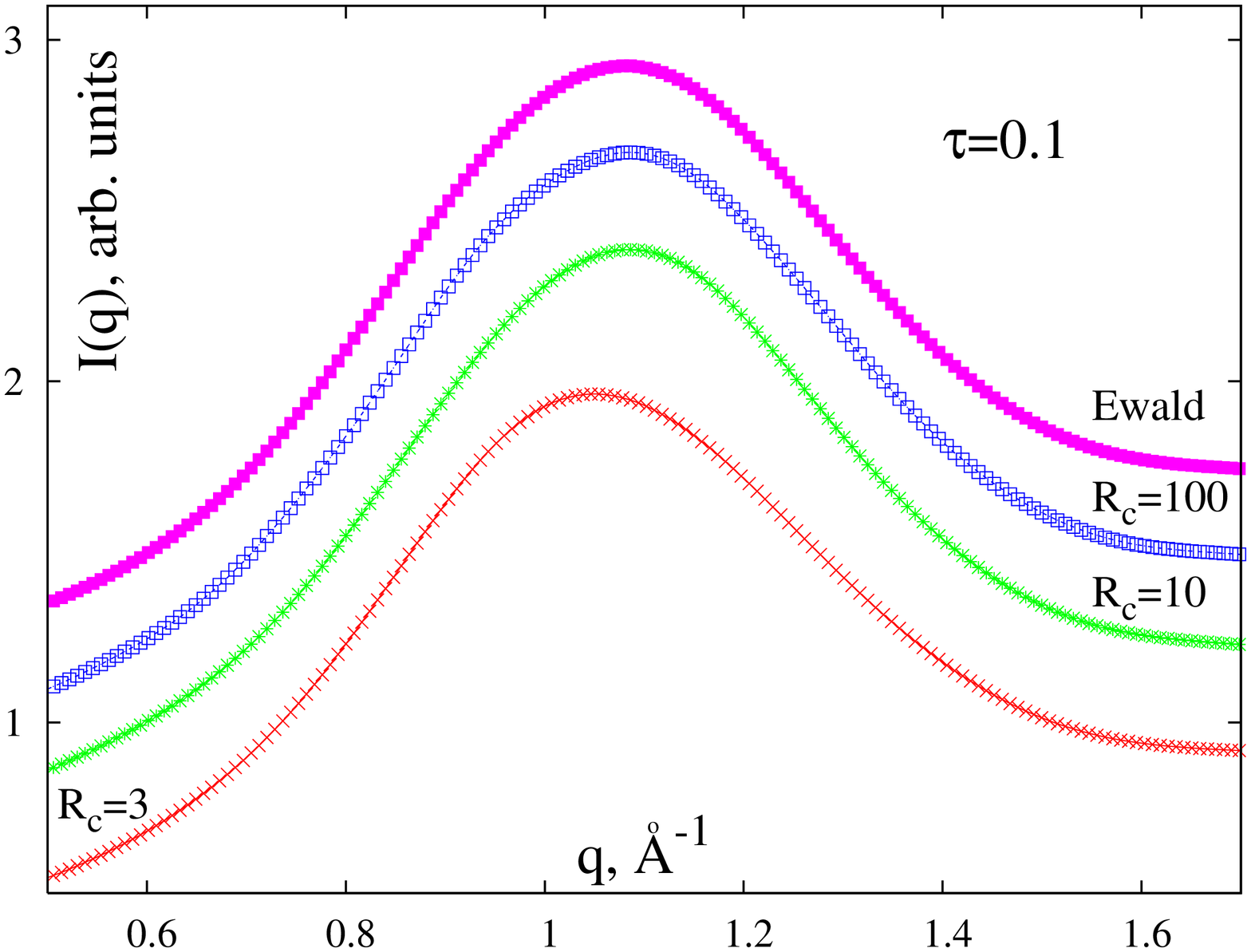}}
\put(215,320){\includegraphics[width=56mm,angle=-90]{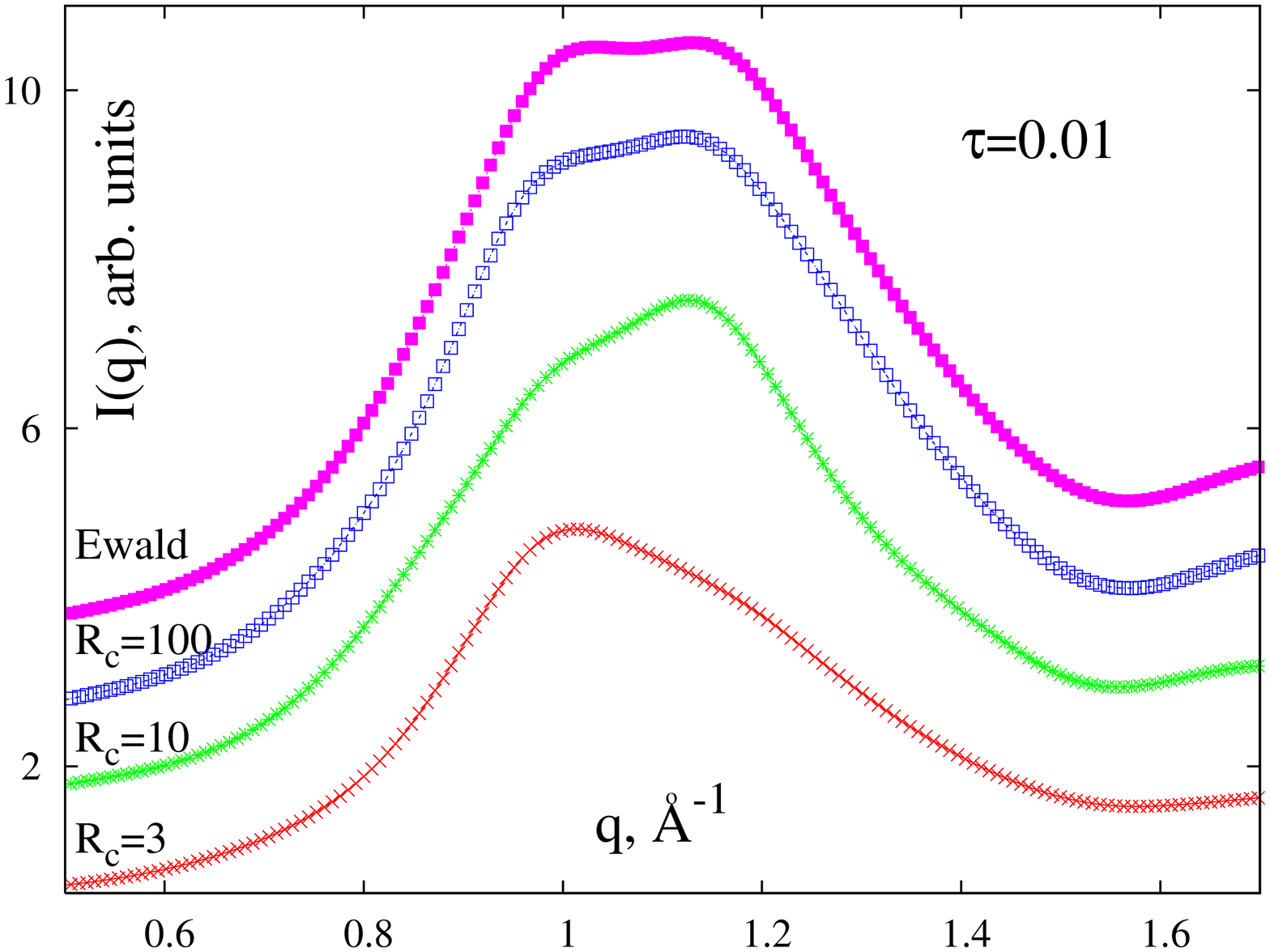}}
\put(5,160){\includegraphics[width=56mm,angle=-90]{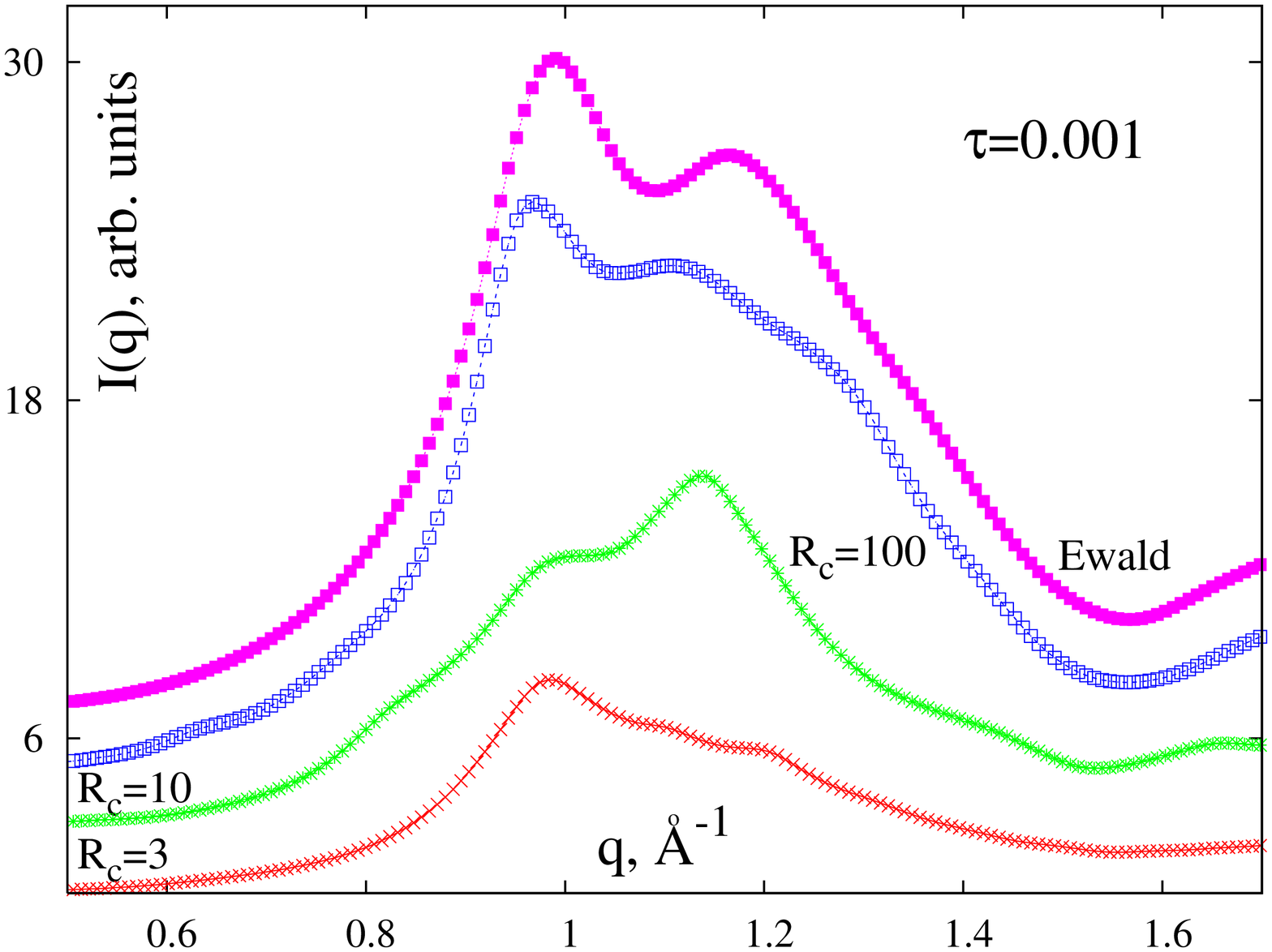}}
\put(215,160){\includegraphics[width=56mm,angle=-90]{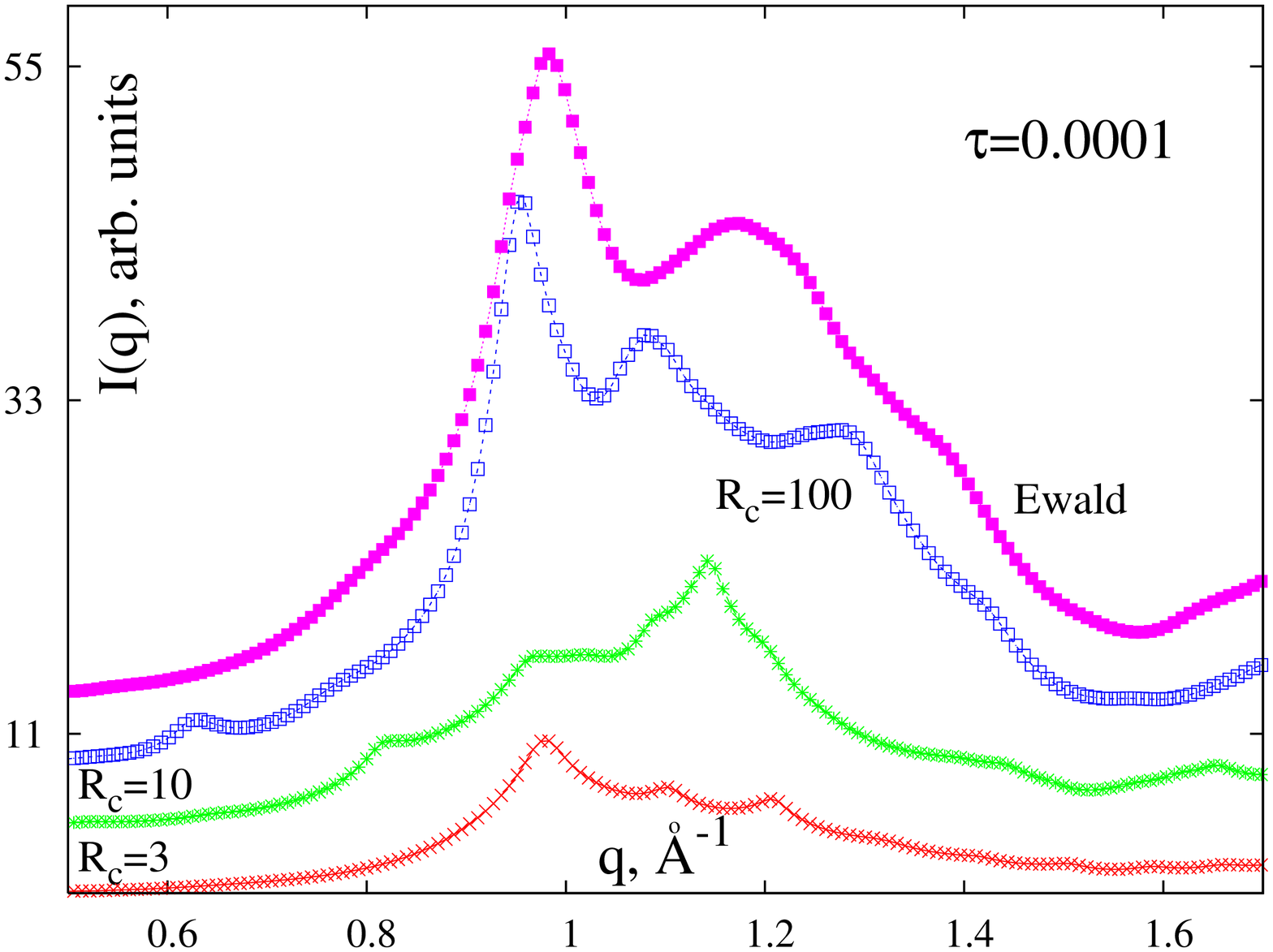}}
\end{picture}
\caption{\label{fc.eps}
Theoretical powder neutron scattering profiles 
of model (\ref{Ham}) at the dipolar cut-off distances 
$R_c=3,10,100,\infty$ and the dimensionless temperatures 
$\tau=0.1,\,0.01,\,0.001,\,0.0001$.
At $\tau=0.1$ the profiles are not sensitive to the cut-off distance
and reflect in fact the properties of the nearest-neighbor 
Heisenberg model without dipolar interactions. 
With the approach to the critical temperature the profiles acquire dispersions
that are unique reflections of different ordering tendencies of 
model (\ref{Ham}) at different $R_c$. The profiles are uniformly off-set for clarity.
}
\end{figure*}

We now proceed to calculate the neutron scattering intensity $I({\bf q})$ 
of model (\ref{Ham}) within the MFT scheme, Eq.~(\ref{IqMFT}).
Unlike the spectra $\lambda^{\alpha}({\bf q})$, 
$I({\bf q})$ can be directly compared to experiments, such as the one
on a powder sample of GGG in zero external magnetic field
\cite{Petrenko-PRL}.
We determine the powder intensity $I(q)$ by numerically calculating
the spherical average of $I({\bf q})$, 
which entails the application of 
a cubic interpolation procedure separately to the
numerator and denominator of Eq.~(\ref{IqMFT}). 
To illustrate the influence of $R_c$ on the spin-spin correlations of model (\ref{Ham}),
we show in Fig.~\ref{fc.eps}
the MFT powder scattering profiles $I(q)$ of model (\ref{Ham}) at several
$R_c=3,10,100,\infty$. 

The $I(q)$ profiles show
different degrees of dependence on the dipolar cut-off distance $R_c$ 
at different dimensionless MFT temperatures $\tau$ (Fig.~\ref{fc.eps}).
At temperatures sufficiently far from the mean-field critical regime 
(Fig.~\ref{fc.eps}, $\tau =0.1$),
the magnetic correlations are not very sensitive to $R_c$.
In fact, it has been found in Monte Carlo simulations~\cite{Petrenko-PRL},
that paramagnetic liquid-like correlations of GGG can be well described
by completely ignoring the dipolar term.
At $\tau\approx0.01$ the profiles start to 
capture the effect of the dipolar interactions on the magnetic correlations.
The effect turns out to 
depend on $R_c$.
Indeed, at $\tau\approx0.001$\footnote
{The powder MFT Bragg intensities grow as $\log|\tau|$
as opposed to the $1/\tau$ growth of the ${\bf q}$-dependent intensities. 
This explains the necessity of considering rather small $\tau$ in order to theoretically
approach the critical regime.}
the profiles clearly display a specific
${q}-$dependence sensitive to the chosen $R_c$.
Starting from this regime, the influence
of small $R_c$ becomes uncontrollable, as seen by the formation of 
Bragg peaks at spurious ordering wave vectors (cf. Fig.~\ref{hhlRc.eps}).
Figure~\ref{fc.eps} 
shows that even the consideration of 
a rather large dipolar cut-off of $R_c=100$ does
not allow one to reproduce magnetic correlations consistent with the 
physical $R_c=\infty$ limit.
This, together with the incommensurability
of the fundamental ordering wave vector of the physical $R_c=\infty$ case: 
${\bf q}_{\rm ord}= 2\pi/a \; (0.348\;0.348\;0.253)$ strongly warns against 
using a standard Monte Carlo method with periodic boundary conditions to tackle this problem.
Moreover, 
we anticipate that a $1/L^3$ finite-size
correction of the real space representation of the
Ewald interactions would be sufficiently large
for numerically accessible system sizes 
so as to prohibit a quantitative disentanglement of the role of 
perturbative terms~\cite{Petrenko-PRL,condmat0511403,Rimai} to model (\ref{Ham}),
presumably a necessary condition
for obtaining a quantitative description of the experimental 
incommensurate magnetic correlations~\cite{Petrenko-PRL} in GGG.

\section{Conclusion}
To conclude, we have identified another example of a highly frustrated 
Heisenberg antiferromagnetic 
system, namely that on the garnet lattice, where the 
selection of the soft mode is sensitive to an ad-hoc cut-off distance $R_c$ 
of the dipolar interactions.
This adds to the cases of the Ising 
(spin ice)~\cite{Gingras-CJP,Melko,Isakov-PRL} and the dipolar Heisenberg
antiferromagnets~\cite{Enjalran-condmat,Cepas}, 
both on the pyrochlore lattice.
Continued progress in understanding GGG 
may be 
possible provided the dipolar interactions
are treated at their physical infinite cut-off limit $R_c=\infty$.
This will, however, 
require systematic 
investigations of the roles of long-range exchange \cite{condmat0511403}
and single ion anisotropy~\cite{Rimai} 
in this material.
Finally, we note that the positions and relative intensity of the $I(q)$
maxima differ dramatically from those found experimentally 
(Fig. 2a of Ref.~\cite{Petrenko-PRL});
an adjustment of $\tau$ does not solve the discrepancy.
This may indicate that a quantitative description of the low-temperature
spin-spin correlations
in GGG requires inclusion of exchange
interactions beyond nearest neighbours \cite{Wolf,condmat0511403} and/or single ion
anisotropy.

\ack
Support for this work was provided by the NSERC of
Canada, the Canada Research Chair Program (Tier I,
M.G), the Province of Ontario and the Canadian Institute 
for Advanced Research. M.G. thanks the U. of Canterbury 
(UC) for an Erskine Fellowship and the hospitality of 
the Department of Physics and Astronomy at UC 
where part of this work was completed.

\end{document}